\documentclass[preprint,superscriptaddress,nofootinbib]{revtex4}
\usepackage[dvips]{graphicx}
\usepackage[usenames]{color}
\usepackage{setspace}
\usepackage{amsmath}
\usepackage{amssymb}
\usepackage{amsthm}
\usepackage{epsf}
\usepackage{epsfig}

\begin{document}
\preprint{CTPU-PTC-18-09}

\newcommand{\GeV}{{\text{GeV}}}
\newcommand{\Hz}{{\text{Hz}}}
\newcommand{\eff}{{\text{eff}}}
\newcommand{\SM}{{\text{SM}}}
\newcommand{\BSM}{{\text{BSM}}}
\renewcommand{\topfraction}{0.8}
\newcommand{\Zhuoni}[1]{{\bf\color{magenta} ZQ: #1}}
\newcommand{\tc}{\textcolor{blue}}

\title{Exploring dynamical CP violation induced baryogenesis by gravitational waves and colliders}

\author{Fa Peng Huang}
\affiliation{Center for Theoretical Physics of the Universe, Institute for Basic Science (IBS),
		Daejeon 34126, Korea}
\author{Zhuoni Qian}
\affiliation{Center for Theoretical Physics of the Universe, Institute for Basic Science (IBS),
		Daejeon 34126, Korea}	
\author{Mengchao Zhang}
\affiliation{Center for Theoretical Physics of the Universe, Institute for Basic Science (IBS),
		Daejeon 34126, Korea}

\begin{abstract}
By assuming a dynamical source of CP violation,
the tension between sufficient CP violation for successful electroweak baryogenesis and strong constraints from current electric dipole moment measurements could be alleviated. We study how to explore such scenarios through gravitational wave detection, collider experiments, and their possible synergies with a well-studied example.

\end{abstract}


\maketitle

\section{Introduction}
\label{sec:introduction}
Electroweak (EW) baryogenesis becomes a promising and testable mechanism at both particle colliders and  gravitational wave (GW) detectors to explain the observed  baryon asymmetry of the Universe (BAU), especially after the discovery of the 125 GeV Higgs boson at the LHC~\cite{Aad:2012tfa,Chatrchyan:2012xdj} and the first detection of GWs by Advanced LIGO~\cite{Abbott:2016blz}.
The long-standing puzzle of BAU in particle cosmology
is quantified by the baryon-to-photon ratio $\eta_B=n_B/n_{\gamma}=5.8-6.6\times 10 ^{-10}$~\cite{Patrignani:2016xqp}
 at $95\%$ confidence level (C.L.),
which is determined from the data of the cosmic microwave background radiation
or the big bang nucleosynthesis.
It is well known that to generate the observed BAU,
Sakharov's three conditions (baryon number violation, $C$ and $CP$ violation, and departure from thermal equilibrium or $CPT$ violation)~\cite{Sakharov:1967dj} need to be satisfied,
and various baryogenesis mechanisms have been proposed~\cite{Dine:2003ax}.
Among them, EW baryogenesis~\cite{Kuzmin:1985mm,Trodden:1998ym,Morrissey:2012db} may potentially relate the nature of the Higgs boson and phase transition GWs.
An important ingredient for successful EW baryogenesis
is the existence of a strong first-order phase transition (SFOPT) which can achieve departure from thermal
equilibrium. The lattice simulation shows that the
$125$ GeV Higgs boson is too heavy for  an efficient SFOPT~\cite{Morrissey:2012db}, nevertheless, there exist already in the literature four types of extensions of the standard model (SM) Higgs sector to
produce a SFOPT~\cite{Chung:2012vg}.
Another important ingredient is sufficient source of $CP$ violation, which is too weak in the  SM.
One needs to introduce a large enough $CP$ violation, which also needs to
escape the severe constraints from the electric dipole moment (EDM) measurement.

Thus,  in this work,
we study the dynamic source of $CP$ violation\footnote{In recent years,  inspiring works on the dynamical $CP$ violation appeared in Refs.~\cite{Baldes:2016rqn,Baldes:2016gaf,Bruggisser:2017lhc,Bruggisser:2018mus}.},  which depends on the cosmological evolution
of a scalar field.
For example, this can be realized  by the two-step
phase transition, where a sufficient $CP$ violation and SFOPT can be satisfied simultaneously to
make the EW baryogenesis work. The studied scenario could explain the observed BAU
while satisfying all the constraints from EDM measurement and collider data.

As a well-studied example,  the SM is extended
with a real scalar field $S$ and a dimension-five operator
 $y_t \frac{\eta}{\Lambda} S \bar{Q}_L \tilde{\Phi} t_R+\rm H.c.$
to provide the SFOPT and sufficient $CP$ violation for EW baryogenesis, which was firstly proposed in Refs.~\cite{Espinosa:2011eu,Cline:2012hg}.
This dimension-five operator actually appears in many composite models
and this source of $CP$ violation for BAU evolves with the scalar field $S$.
At the very early universe, $\langle S \rangle=0$ \footnote{In this work,  we use the angle brackets $\langle \rangle$ to denote the vacuum expectation value  (VEV) of the corresponding field.}, then the value evolves to $\langle S \rangle = \sigma$ through a second-order phase transition.
The CP violating top quark Yukawa coupling is then obtained and can source the BAU in the following SFOPT~\cite{Basler:2017uxn}.
After that, $\langle S \rangle $ evolves to zero again, and the $CP$ violating top quark Yukawa coupling vanishes at tree level.
This evolution of the coupling naturally avoids the strong constraints from the EDM measurements,
and yields distinctive signals at hadron colliders and  lepton colliders,
such as the LHC,  the Circular Electron-Positron Collider (CEPC)~\cite{CEPC-SPPCStudyGroup:2015csa} , the International Linear Collider (ILC)~\cite{Gomez-Ceballos:2013zzn},
 and the Future Circular Collider (FCC-ee)~\cite{dEnterria:2016sca}.
We discuss the constraints on the parameters of the effective Lagrangian from both particle physics experiments and cosmology,
since probing the nature of the EW phase transition or EW baryogenesis is one important scientific goal for fundamental physics after the discovery of the Higgs boson~\cite{CEPC-SPPCStudyGroup:2015csa, Arkani-Hamed:2015vfh}.
This EW baryogenesis scenario with dynamical $CP$ violation should and could be probed by future colliders and help us to
unravel the nature of the Higgs potential and the dynamics of EW symmetry breaking~\cite{CEPC-SPPCStudyGroup:2015csa, Arkani-Hamed:2015vfh}. Especially, the collider signals when we include the dynamical source of $CP$ violation are quite distinctive from the collider signals when only the SFOPT is considered~\cite{Curtin:2014jma}.

After the first discovery of GWs by Advanced LIGO~\cite{Abbott:2016blz},
GWs becomes a new realistic approach to study the EW baryogenesis mechanism by future space-based experiments, such as the approved Laser Interferometer Space Antenna (LISA)~\cite{Seoane:2013qna} (which is assumed
to be launched in 2034),
Deci-hertz Interferometer Gravitational wave Observatory (DECIGO)~\cite{Kawamura:2011zz},
Ultimate-DECIGO (U-DECIGO)~\cite{Kudoh:2005as}, and Big Bang Observer (BBO)~\cite{Corbin:2005ny}.
The SFOPT process in the  EW baryogenesis can produce detectable GW signals through
three mechanisms: bubble collisions, turbulence, and sound waves
\cite{Witten:1984rs,Hogan:1984hx,Turner:1990rc,Kamionkowski:1993fg,Huber:2008hg,Caprini:2009yp,Espinosa:2010hh,No:2011fi,Hindmarsh:2013xza,Hindmarsh:2015qta,Caprini:2015zlo}.

Thus, after considering the GW signals from SFOPT,
we report on a joint analysis of observational signatures from the EW baryogenesis under our scenario, correlating the GW and collider signals.
This type of two-step phase transition with its GW signals, and the EW baryogenesis in this scenario were well-studied in the previous study.
In this work, we recalculate and describe this scenario from the dynamical $CP$ violation perspective  and first investigate
how to explore this scenario by collider signals and their correlations with the GW signals.
The structure of the paper is as follows: in Sec. \ref{sec:Model}, we describe the effective model of the dynamical $CP$ violation for successful baryogenesis.
In Sec.III, we discuss the dynamics of the phase transition in detail.
In Sec.IV,  size of the dynamical CP violation and the BAU are estimated.
In Sec.V, the constraints and predictions from the EDM measurements and colliders are given.
In Sec.VI, we investigate the GW signal and its correlation to the collider signals.
Finally, we conclude in Sec.VII.

\section{cosmological evolution of the Yukawa coupling and baryogenesis}\label{sec:Model}

Based on the fact that sufficient source of $CP$ violation for successful baryogenesis are typically severely constrained
by EDM measurement, there is a possibility that the
$CP$ violating coupling depends on the cosmological evolution history.
During the early Universe, there exists a large $CP$ violation for successful
baryogenesis. When the universe evolves to the current time,
the source of $CP$ violation evolves to zero at tree level.
In this work, we study the $CP$-violating Yukawa coupling which evolves from a sufficiently large value to a loop-suppressed small value at the current time, by assuming
it depends on a dynamical scalar field; i.e.,
the phase transition process can make the $CP-$violating Yukawa coupling transit from a large value to zero at the tree level.
A well-studied example is  the $CP$-violating top Yukawa coupling  scenario as proposed  in Refs.~\cite{Espinosa:2011eu,Cline:2012hg}.
Namely, there exist extra terms to the SM top-quark Yukawa coupling which reads:
\begin{equation}
y_t \eta \frac{S^n}{\Lambda^n} \bar{Q}_L \tilde{\Phi} t_R+h.c.
\end{equation}
where  $y_t=\sqrt{2} m_t/v$ is the SM top-quark Yukawa coupling,  $\eta=a+ i b$ is a complex parameter,
$\Lambda$ is the new physics (NP) scale,  $\Phi$ is the  SM Higgs doublet field,
$Q_L$ is the third-generation $SU(2)_L$ quark doublet, $t_R$ is the right-handed
top quark, and $S$ is a real singlet scalar field beyond the SM.
During the phase transition process in the early universe,
the scalar field $S$ acquires a VEV $\sigma$,  and then a sizable $CP$-violating
top-Yukawa coupling can be obtained and contribute to the EW baryogenesis for BAU.
After the phase transition finishes, the
VEV of S vanishes and the Higgs field acquires a VEV $v$, meaning that the $CP$-violating top-quark
Yukawa coupling vanishes at the tree-level and evades the strong EDM constraints naturally.
More generally,
we can assume that the top-quark Yukawa coupling depends on a scalar field or its VEV, which changes during the cosmological evolution.
For the phase transition case,  the $CP$-violating top-Yukawa coupling simply depends on the phase transition dynamics.

We take the  $n=1$ as a simple but representative example to show how it gives successful
baryogenesis and how it is detected with the interplay of collider experiments and gravitational wave detectors.
The corresponding effective Lagrangian~\cite{Espinosa:2011eu,Cline:2012hg,Huang:2015bta} can be written as:
\begin{equation}\label{lag}
{\cal L}={\cal L}_{\rm SM}
- y_t \frac{\eta}{\Lambda} S \bar{Q}_L \tilde{\Phi} t_R+\rm H.c+\frac{1}{2} \partial_\mu S \partial^\mu S  + \frac{1}{2}\mu^2  S^2
- \frac{1}{4}\lambda S^4  - \frac{1}{2} \kappa S^2 (\Phi^\dagger \Phi) .
\end{equation}
Based on this Lagrangian, we study the collider constraints, predictions, GW signals and EDM constraints in detail.
For simplicity, we choose the default values as $a=b=1$,
namely, $\eta=1+i$.
We can, of course rescale $\eta$ and $\Lambda$ simultaneously to keep the effective field theory
valid up to the interested energy scales.
It is not necessary to consider the domain wall problem here as shown in Refs.~\cite{Espinosa:2011eu,Barger:2011vm}.
The coefficients $\mu^2$, $\lambda$, and $\kappa$ are assumed to be positive in this work.
It worth noticing that we just use the same Lagrangian in Refs.~\cite{Espinosa:2011eu, Cline:2012hg}
to realize the two-step phase transition and do not consider other  possible operators, which may make the two-step
phase transition difficult to  realize. If we neglect the dimension-five operator, there is a $Z_2$ symmetry in the potential,  which
makes the two-step phase transition more available.

For the above  effective Lagrangian, a second-order and first-order phase transition could occur in orders.
First, a second-order phase transition happens, the scalar field $S$ acquires a VEV, and
the dimension-five operator generates a sizable $CP$-violating top-Yukawa coupling,
which provides the source of $CP$ violation needed for BAU.
Second, a SFOPT occurs when the vacuum transits from $(0,\langle S \rangle)$ to $(\langle \Phi \rangle,0)$.
After the two-step phase transition,\footnote{There are extensive studies on the two-step phase transition in the models of an extended Higgs sector with singlet scalars as in Refs.~\cite{Huang:2016cjm,Hashino:2016xoj,Vaskonen:2016yiu,
Beniwal:2017eik,Cline:2017qpe,Kurup:2017dzf,Chao:2017oux,Jiang:2015cwa,
Baker:2017zwx,Demidov:2017lzf,Wan:2018udw}.} the VEV of $S$ vanishes at the tree level, which naturally avoids the electron and neutron EDM constraints, and the dimension-five operator induces the interaction term $-\frac{ m_t}{\Lambda}(a S \bar{t} t+ib S \bar{t}\gamma_5 t)$,
which produces abundant collider phenomenology at the LHC and future lepton colliders, such as CEPC, ILC, and FCC-ee.

It is worth noticing that the dimension-five
effective operator $y_t \frac{\eta}{\Lambda} S \bar{Q}_L \tilde{\Phi} t_R$ is present as well in  some 
NP models~\cite{Cline:2017jvp,Gripaios:2009pe,Chala:2017sjk},   especially many
composite Higgs models~\cite{Gripaios:2009pe,Chala:2017sjk}.
For example, the singlet and the dimension-five operator
can come from composite Higgs models such as
$SO(6)\times U(1)^{\prime}/SO(5) \times U(1)^{\prime}$,
 $SO(5)\times U(1)_S \times U(1)^{\prime}/SO(5)\times U(1)^{\prime}$,
or  $SO(6)\to SO(5)$~\cite{Gripaios:2009pe,Chala:2017sjk}.

\section{Phase transition dynamics}\label{sec:EWPT}

In this section we discuss the phase transition dynamics, which provides the necessary conditions for
EW baryogenesis and produces detectable GWs during a SFOPT.
To study phase transition dynamics, we use the
the methods in Refs.~\cite{Dolan:1973qd,Carrington:1991hz,Quiros:1999jp}
and write the effective potential as a function of spatially homogeneous background scalar fields, i.e.,  $ S(x) \to \sigma(x)$
and $ \Phi(x)\to \frac{1}{\sqrt{2}} (0, H(x))^T$.
Thus, the full finite-temperature
effective potential up to the one-loop level can be written as
\begin{equation}\label{fullpotential}
V_{\rm eff}(H,\sigma,T)=V_\text{tree}(H,\sigma)+\Delta V_1^{T\neq 0}(H,\sigma,T)  +V_1^{T=0}(H,\sigma)    \,\,,
\end{equation}
where $V_\text{tree}(H,\sigma)$ is the tree-level potential at zero temperature as defined below in Eq.(\ref{treev}), $\Delta V_1^{T\neq 0}(H,\sigma,T)$ is the one-loop thermal corrections including the daisy resummation, and $V_1^{T=0}(H,\sigma)$ is the Coleman-Weinberg
potential at zero temperature.

The tree-level potential at zero temperature in Eq.~(\ref{fullpotential}) is
\begin{equation}\label{treev}
V_\text{tree}(H,\sigma)=-\frac{1}{2}\mu_{SM}^2 H^2-\frac{1}{2}\mu^2 \sigma^2+\frac{1}{4}\lambda_{SM} H^4+\frac{1}{4}\lambda \sigma^4+\frac{1}{4}\kappa H^2\sigma^2.
\end{equation}
We can see that there are four distinct extremal points\footnote{Actually, there are nine extremal points. However, we do not consider the negative $H$ or $\sigma$ in this work.}, and
requiring only two global minima at $V(\mu_{SM}/\sqrt{\lambda_{SM}},0)$ and
$V(0, \mu/\sqrt{\lambda})$ leads to the relation
$\kappa>2 \sqrt{\lambda \lambda_{SM}}$.
When $\frac{\mu_{SM}^{4}}{ \lambda_{SM}}=\frac{\mu^{4}}{ \lambda}$,
the two minima at tree level degenerate, and if
 $\frac{\mu_{SM}^{4}}{ \lambda_{SM}}>\frac{\mu^{4}}{ \lambda}$,
$V(\frac{\mu_{SM}}{\sqrt{\lambda_{SM}}},0)$ becomes the only global minimum.
The SFOPT
can be realized easily since the potential barrier height appears at tree level
and is not suppressed by loops or thermal factors.
Based on these properties,  it is convenient to parameterize $\lambda$ and $\mu^2$ as
\begin{eqnarray}
  \lambda =(\frac{\kappa}{2 \lambda_{SM}})^2 \lambda_{SM}(1+\delta_\lambda),~~
   \mu^2 = \mu^2_{SM}\frac{\kappa}{2\lambda_{SM}} (1+\delta_{\mu^2}).
\end{eqnarray}
Later on we use the full effective potential in  Eq.(\ref{fullpotential})  to numerically calculate the
phase transition dynamics and GW signal,
but first we can qualitatively understand the phase transition dynamics
using the tree-level potential and leading-order temperature correction,
since the full one-loop effective potential only sightly
modifies the values of the parameter space.
Thus, using the high-temperature expansion up to leading order $\mathcal{O}(T^2)$,  the effective thermal potential
in Eq.(\ref{fullpotential}) can be approximated as
\begin{equation}\label{eq:highT_V}
 V(H,\sigma,T)=(D_HT^2-\frac{\mu^2_{SM}}{2})H^2+(D_\sigma T^2-\frac{\mu^2}{2})\sigma^2
+\frac{1}{4}(\lambda_{SM} H^4+ \kappa H^2\sigma^2 + \lambda \sigma^4)
\end{equation} with
\begin{align*}
D_H=\frac{1}{32}(8\lambda_{SM}+g'^2+3 g^2+4 y_t^2+2 \kappa/3),~~
D_\sigma =\frac{1}{24}(2\kappa+3 \lambda) \,\,     ,
\end{align*}
where the SM $U(1)$  gauge coupling  $g'=0.34972$, $SU(2)$  gauge coupling  $g=0.65294$, and  top-quark Yukawa $y_t=0.99561$~\cite{Buttazzo:2013uya}.
The terms $D_H T^2$ and $D_{\sigma} T^2$ represent  the leading-order thermal corrections to the  fields  of $H$
and $\sigma$, respectively.  Here,  the contributions from the dimension-five
operator are omitted as similarly argued and dealt with in Refs.~\cite{Espinosa:2011eu,Cline:2012hg,Huang:2015bta}.
Thus,  the washout parameter can be approximated  as
\begin{equation}\label{vtc}
\frac{v(T_c)}{T_c}\sim  \frac{2  v}{m_H }  \sqrt{\frac{D_H-D_{\sigma}}{\delta_\lambda-2\delta_{\mu^2}}}  \,\,\,\,.
\end{equation}
Numerically, the  allowed parameter space for large washout parameter $v(T_c)/T_c$  is shown in Fig.\ref{vtc} for
$\kappa=1.0$ and $\kappa=2.0$ cases, respectively.
\begin{figure}[h] 
   \centering
   \includegraphics[width=5.26in]{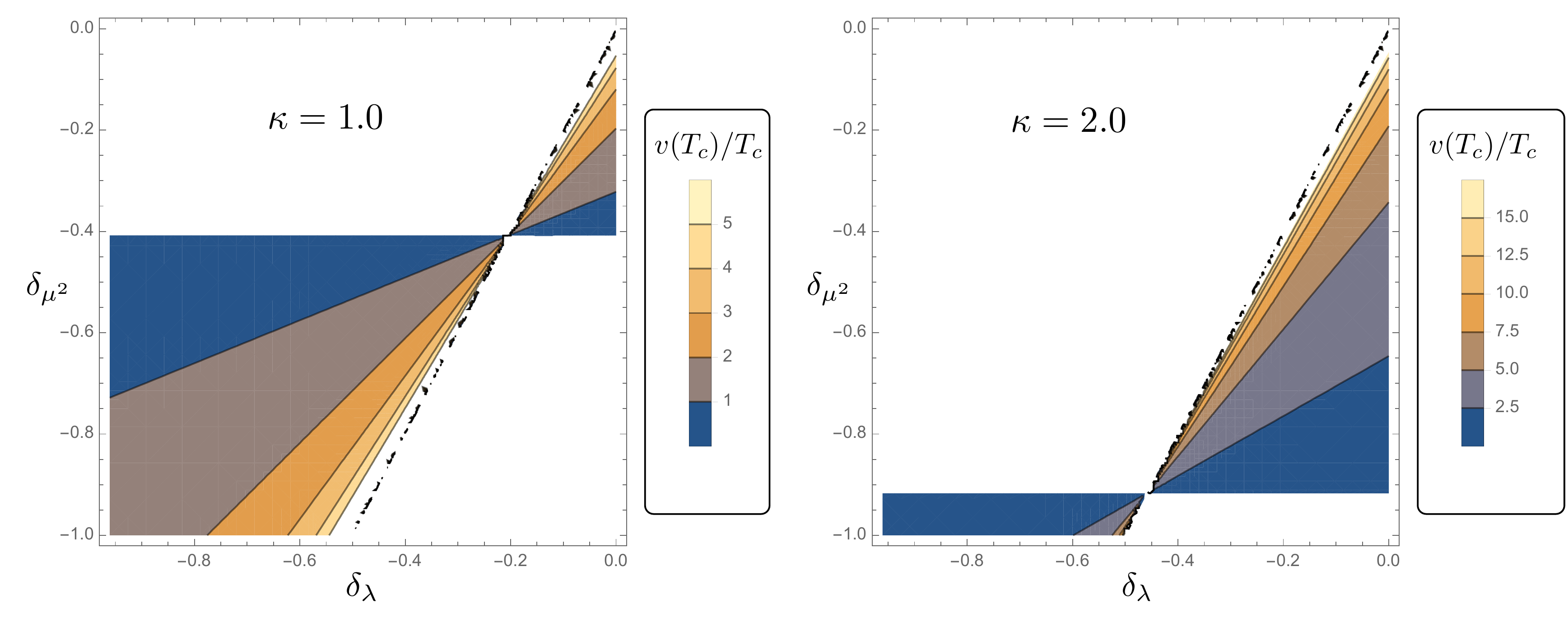}
      \caption{Parameter spaces for a large washout parameter for $\kappa=1.0$ and $\kappa=2.0$, respectively.}
   \label{vtc}
\end{figure}
We use the washout parameter to qualitatively see the SFOPT-favored parameter region.
Generally speaking, larger washout parameter represents a stronger first-order phase transition.
For the quantitative determination of the SFOPT, we need to calculate the nucleation temperature $T_N$
as discussed below. Eventually, some typical parameter sets that give a two-step phase transition (the phase transitions take place as $(0,0) \to (0,\langle S \rangle) \to (\langle \Phi \rangle,0)$ with the decreasing of the
temperature, where SFOPT occurs during the second step) and
produce a SFOPT  are shown in
Table.~\ref{typical}.
\begin{table}[h!]
\centering
\caption{Some typical parameter  points, which can give  a two-step phase transition and SFOPT.}
\begin{tabular}{ccccccc}
  \hline
  \hline
  $\kappa$ & $\delta_{\lambda}$ & $\delta_{\mu^2}$  & $T_N$[GeV]  \\

  \hline
  0.88     & -0.21              & -0.61                   & 128.4     \\

  \hline
  0.88     & -0.21              & -0.51                & 171.8        \\
   \hline
  0.88     & -0.21              & -0.41                  & 115.3       \\     \hline
  1.00    & -0.21              & -0.41                  & 116.0      \\  \hline
   2.00    & -0.21              & -0.41              & 121.1   \\  \hline
   2.00    &-0.21               &-0.22              & 106.6     \\ \hline
   2.00    & -0.21              &-0.30                      &  113.6     \\   \hline
     4.00    & -0.21              & -0.21                 & 115.9  \\  \hline   \hline
\end{tabular}
\label{typical}
\end{table}

We now describe  the methods used to obtain the values in Table.~\ref{typical}. 
We first introduce two important quantities $\alpha$ and $\tilde{\beta}$,
which can precisely describe the dynamical properties of the phase transition~\cite{Grojean:2006bp}.
The key quantity to obtain $\alpha$ and $\tilde{\beta}$
is the bubble nucleation rate per unit volume per unit time
$  \Gamma=\Gamma_0 \exp[-S_E]$,
where $S_E(T) = S_3(T)/T$ is the Euclidean action
and $\Gamma_0 \propto T^4$.
And $S_3$ is the three-dimensional Euclidean action, which can be expressed as
\begin{equation}\label{eq:actionfunc}
S_3= \int d^3\,r \left\{\frac{1}{2}\left(\frac{d H}{dr}\right)^2+\frac{1}{2}\left(\frac{d \sigma}{dr}\right)^2+V_{\textrm{eff}}(H,\sigma,T)\right\}.
\end{equation}
To calculate the nucleation rate,  we need to obtain the profiles of the two scalar fields.
Here, we need to deal with phase transition dynamics involving two fields using the method in
 Refs.~\cite{Cline:1999wi,Profumo:2010kp,Wainwright:2011kj}
 by choosing a path $\vec{\varphi}(t)=\left(H(t),\sigma(t)\right)$ that connects the initial and final vacuum.
 Then, we can get the bounce solution $\varphi_b$ by solving the following differential bounce equation
\begin{align}
  \frac{d^2 \varphi_b^{}}{d r^2}
  + \frac{2}{r} \frac{d \varphi_b^{}}{d r}
  =\frac{\partial V_{\rm eff}^{}}{\partial \varphi_b^{}} ,
\end{align}
with the boundary conditions
\begin{align}
  \lim_{r \to \infty} \varphi_b = 0,
  \quad
  \left. \frac{d \varphi_b^{}}{d r} \right|_{r=0} = 0.
\end{align}
After we obtain the nucleation rate, the parameter $\tilde{\beta}$ can be defined as
\begin{align}
  \tilde{\beta} =
  T_N\frac{d}{d T}\left(\frac{S_3(T)}{T}\right)\Bigg|_{T=T_N }.
\end{align}
Another important quantity $\alpha$ parametrizes the ratio between the false-vacuum energy
density $\varepsilon(T_N^{})$ and the thermal energy density
$\rho_{\rm rad}^{}(T_N^{})$ in the symmetric phase at the nucleation
temperature $T_N$. It is defined as
\begin{align}
  \alpha = \frac{\varepsilon(T_N)}{\rho_{\rm rad}(T_N)} ,
\end{align}
where the thermal energy density is given by
$\rho_{\rm rad}^{}(T) =(\pi^2/30) g_*^{}(T)T^4$
with $g_*^{}$ being the relativistic degrees of
freedom in the thermal plasma.
And $\varepsilon(T)$ is defined as
\begin{align}
 \varepsilon(T)
  = - V_{\eff}(\varphi_B(T),T)+T
  \frac{\partial V_{\eff}(\varphi_B(T), T)}{\partial T},
\end{align}
where $\varphi_B(T)$ is the VEV of the broken phase minimum at temperature $T$.

To calculate the parameters $\alpha$ and $\tilde{\beta}$, it is necessary to determine the
nucleation temperature $T_N$ where
the nucleation rate per Hubble volume per Hubble time reaches unity as $\Gamma/\mathcal{H}^4 |_{T=T_N} \simeq 1$,
where $\mathcal{H}$ is the Hubble parameter.
Thus the condition can be simplified as
\begin{align}
  \frac{S_3(T_N^{})}{T_N^{}} =4\ln (T_N/100 \rm GeV)+ 137.
\end{align}

Using the method above, we are able to numerically calculate $T_N$, $\alpha$, and $\tilde{\beta}$.
For the following discussion, we pick two benchmark sets which can produce a two-step phase transition and SFOPT,
and the parameters $\alpha$, $\tilde{\beta}$, $m_S$, and $T_N$ are listed in Table~\ref{ben}.
Usually, a larger $\alpha$ and smaller $\tilde{\beta}$ give a stronger first-order phase transition and
stronger GWs.

\begin{table}[h!]
\centering
\caption{Benchmark points, which can give a SFOPT and  produce phase transition GWs.}
\begin{tabular}{ccccccc}
  \hline
  \hline
Benchmark set & $\kappa$  & $m_S$ [GeV] & $T_N$ [GeV]  & $\alpha$ &  $\tilde{\beta}$   \\
  \hline
I &  2.00              & 115        & 106.6   & 0.035   & 107  \\ \hline
 II&  2.00              &135        &  113.6   & 0.04     & 120   \\   \hline
  \hline
\end{tabular}
\label{ben}
\end{table}

\section{Electroweak Baryogenesis and $CP$ violation}
In this section, we estimate the constraints on the dynamical source of $CP$ violation from the observed value of BAU.
To produce the observed baryon asymmetry from EW baryogenesis,  $CP$ violation is necessary to produce
an excess of left-handed fermions versus right-handed fermions and then generate net baryon excess through EW sphaleron
process~\cite{Espinosa:2011eu,Cline:2012hg}.
After the first step of phase transition, $S$ field obtains a VEV,  and then the $CP$-violating
top-quark Yukawa coupling is obtained.
Thus, during the SFOPT, the top quark  in the bubble wall has a
spatially varying complex mass, which is given by~\cite{Espinosa:2011eu,Cline:2012hg}
$m_t(z) = {y_t\over\sqrt{2}} H(z) \left(1 +(1+ i) {S(z)\over\Lambda}\right)\equiv
		|m_t(z)| e^{i\Theta(z)}$,
where $z$ is the coordinate  perpendicular to the bubble wall.
The $CP$-violating phase $\Theta$ will provide the necessary $CP$ violation for the BAU.
Taking  the transport equations  in Refs.~\cite{Fromme:2006wx,Cline:2011mm,Cline:2012hg,Kobakhidze:2015xlz},
one can estimate the BAU as
\begin{equation}
	\eta_B = {405\Gamma_{\rm sph}\over 4\pi^2 \tilde{v}_{b} g_*T}\int dz\, \mu_{B_L}
	f_{\rm sph}\,e^{-45\, \Gamma_{\rm sph}|z|/(4 \tilde{v}_{b})},
\label{baueq}
\end{equation}
where $\tilde{v}_b$ is the relative velocity between the bubble wall and plasma front in the deflagration case (the bubble wall velocity $v_b$ is smaller than
the sound velocity $c_s =\sqrt{3}/3 \sim 0.57$ in the plasma).
Here,  $\mu_{B_L}$ is the left-handed baryon chemical potential, $\Gamma_{\rm sph}$ is the sphaleron rate, and 
$f_{\rm sph}$  is a function that turns off quickly in the broken phase.
The position-dependent $\Theta(z)$ can provide the CP-violating source in the transport equations and contribute to net left-handed baryon $\mu_{B_L}$.
Here,   we choose $\tilde{v}_{b}\sim 0.2$, which is smaller than the bubble wall velocity $v_b$~\cite{No:2011fi}.
It is because the EW baryogenesis usually favors the deflagration bubble case, and the BAU depends on the
relative velocity between the bubble wall and the plasma front. Thus, we have reasonably small relative velocity
$\tilde{v}_b$, which is favored by the EW baryogenesis to guarantee a sufficient diffusion process in front of the bubble wall
and large enough bubble wall velocity $v_b$ to produce stronger phase transition GWs
(In the deflagration case, a larger bubble wall velocity gives stronger GWs~\cite{Espinosa:2010hh,No:2011fi}).
We take the default value of the bubble wall velocity
$v_b  \sim 0.5$, which is reasonable since the difference between $\tilde{v}_b$ and $v_b$ can be large for a SFOPT with a large washout parameter in the deflagration case.

From the roughly numerical estimation, we see that the observed BAU can be obtained
as long as $\Delta \sigma/\Lambda  \sim 0.1-0.3$, where $\Delta \sigma$ is the change of $\sigma$ during the phase transition and is determined by the phase transition dynamics.
For the two benchmark sets given in Table.~\ref{ben},  the needed $\Lambda$ should be around 1 TeV.
Larger $\Lambda$ gives smaller baryon density, and smaller $\Lambda$ produces an overdensity.
The exact calculation of $\eta_B$ would need improvements from the nonperturbative dynamics of the phase transition
and higher order calculations.
In the following, we discuss how to  explore the parameters from the GWs, EDM data, and collider data,
which offer accurate constraints or predictions.

\section{Constraints and predictions in particle physics experiments}\label{sec:Decay}
After the SM Higgs obtains a VEV $v$ at the end of the SFOPT,  
the SM Higgs doublet field  can be expanded around the VEV as $ \Phi(x)\to \frac{1}{\sqrt{2}} (0, v+H(x))^T$. Thus,
the interaction between $S$ and the top quark becomes
\begin{eqnarray}
\mathcal{L}_{Stt} = - \left(\frac{m_t}{\Lambda} + \frac{m_t H}{\Lambda v} \right) S \left(a\bar{t}t + ib\bar{t}\gamma_5 t   \right).
\end{eqnarray}
Top-quark loop-induced interactions between the scalar $S$ and vector pairs are important in our collider phenomenology study. In this work, $m_S$, $m_H$, and $m_S + m_H$ are all assumed smaller than $2m_t$, and $m_S > m_H/2$. So we can in most cases integrate out top-quark loop effects and use effective couplings to approximately describe the interactions.
Here we use the covariant derivative expansion (CDE) approach~\cite{Gaillard:1985uh,Cheyette:1987qz,Henning:2014wua} to calculate our effective Lagrangian.
After straightforward calculations we obtain the relevant one-loop effective operators
\begin{eqnarray}
\mathcal{L}^{\prime}_{SVV} &=&  \frac{a \alpha_S}{12\pi \Lambda} S G^a_{\mu\nu}G^{a\mu\nu}
-  \frac{b\alpha_S}{8\pi \Lambda} S G^a_{\mu\nu}\tilde{G}^{a\mu\nu} \\\nonumber
&+& \frac{2a\alpha_{EW}}{9\pi \Lambda} S F_{\mu\nu}F^{\mu\nu}
- \frac{b\alpha_{EW}}{3\pi \Lambda} S F_{\mu\nu}\tilde{F}^{\mu\nu}.
\end{eqnarray}
Detailed calculations can be referred in the Appendix.

Another effect that needs to be considered here is the one-loop mixing effect between the particle $S$ and $H$.
In our tree-level Lagrangian, there is no mixing term between the $S$ and $H$, but such a mixing term will be induced by the top-quark loop.
Considering the  one-loop correction, the (squared) mass matrix terms of the scalar fields can be written as
\begin{eqnarray}
\mathcal{L}_{mass} = -\frac{1}{2} \left(\begin{array}{cc}S & H\end{array}\right)
  \left(\begin{array}{cc} {m_S^2}_{\text{,tree}} + \Delta {m_S^2} & \Delta m^2_{HS} \\ \Delta m^2_{HS} &  {m_H^2}_{\text{,tree}} + \Delta {m_H^2}  \end{array}\right)
    \left(\begin{array}{c}S \\H\end{array}\right)    \,\,.
\end{eqnarray}
Those corrections are
\begin{eqnarray}
\Delta {m_H^2} = \frac{3 m_t^4}{4 \pi^2 v^2} \ , \quad
\Delta m^2_{HS} = a\frac{3 m_t^4}{2 \pi^2 \Lambda v} \ , \quad
\Delta {m_S^2} = (a^2-b^2) \frac{3 m_t^4}{4 \pi^2 \Lambda^2}  \ .
\end{eqnarray}
The calculation details  can also be found in the Appendix. This mass matrix can be diagonalized by a rotation matrix $\mathcal{O}$:
\begin{eqnarray}
\mathcal{O}
 \left(\begin{array}{cc} {m_S^2}_{\text{,tree}} + \Delta {m_S^2} & \Delta m^2_{HS} \\ \Delta m^2_{HS} &  {m_H^2}_{\text{,tree}} + \Delta {m_H^2}  \end{array}\right)
 \mathcal{O}^T =
 \left(\begin{array}{cc} m_{S,\text{phy}}^2 & 0 \\ 0 & m_{H,\text{phy}}^2  \end{array}\right) .
\end{eqnarray}
Here $m_{H,\text{phy}} = 125$~GeV is the mass of the SM-like Higgs boson observed by the LHC, and the physical mass eigenstates are the mixing of the scalar fields $H$ and $S$:
\begin{eqnarray}
\left\{\begin{array}{c} S_{\text{phy}} = \mathcal{O}_{11} S + \mathcal{O}_{12} H, \\
 H_{\text{phy}} = \mathcal{O}_{21} S + \mathcal{O}_{22} H. \end{array}\right.  
\end{eqnarray}
From now on we neglect the subscript ``$\text{phy}$'' and all the fields and masses are physical by default.

\subsection{Electric dipole moment experiments}

\begin{figure}[htbp]
\begin{center}
\includegraphics[width=16.0cm]{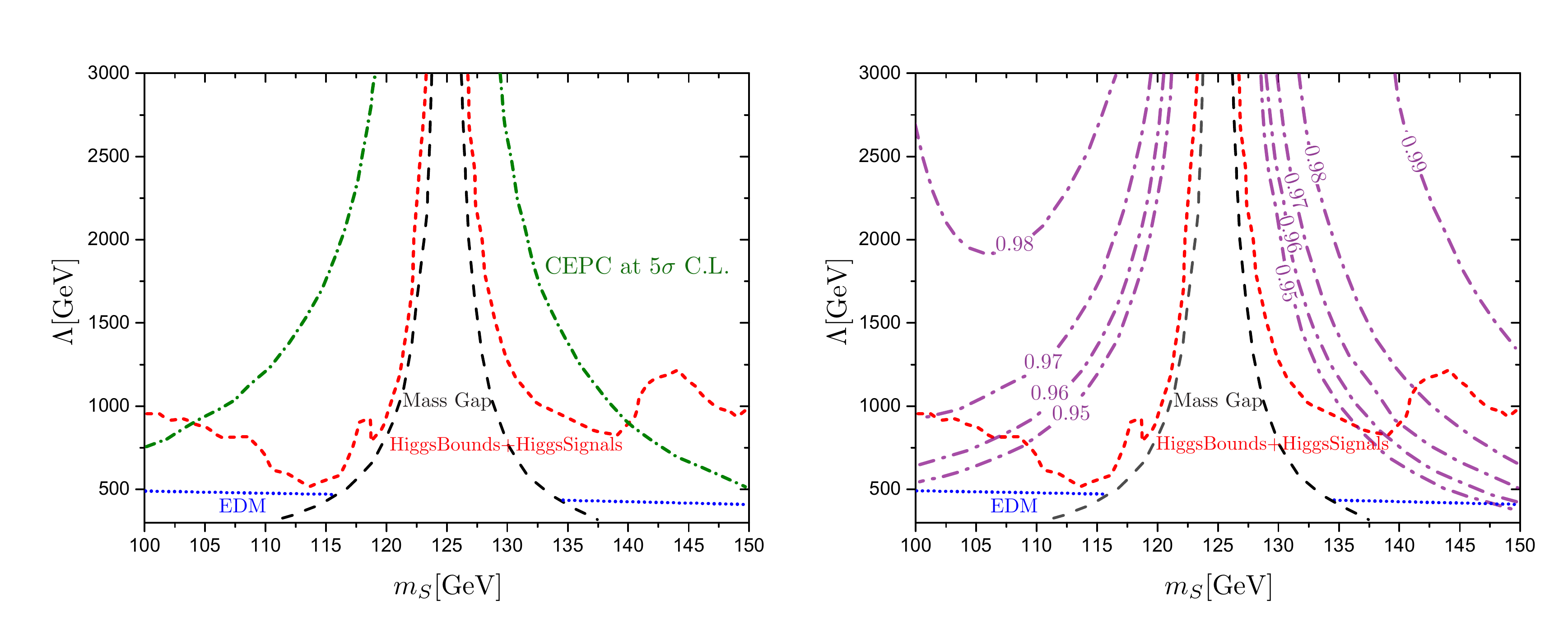}
\caption{
Current exclusion limit and future search sensitivity projected on $\Lambda$ versus $m_S$ plane.
In both plots, regions between black dashed lines are forbidden by mass mixing term $\Delta m^2_{HS} = a\frac{3 m_t^4}{2 \pi^2 \Lambda v}$;
regions below dotted blue lines have been excluded by EDM measurement;
regions below dashed red lines have been excluded by collider scalar searches and Higgs data.
In the left plot, regions below dash dotted olive lines can be observed from $ZS$ production at 5 ${ab}^{-1}$ CEPC with a C.L. higher than 5$\sigma$.
In the right plot, we show the ratio $\frac{\sigma(HZ)}{\sigma_{SM}(HZ)}$ with purple dash dotted contour lines.
In the plot, we set $a = b = 1$ and $\kappa = 2$.}
\label{Limit}
\end{center}
\end{figure}

Current EDM experiments put severe constraints on many baryogenesis models.
For example, the ACME Collaboration's new result, i.e. $\left | d_e \right | < 8.7 \times 10^{-29} \, {\rm cm\cdot e}$ at 90\% C.L.~\cite{Baron:2013eja},
has ruled out  a large  portion of the $CP$ violation parameter space for many baryogenesis models.
However,  in this dynamical $CP$ violation baryogenesis scenario,
the strong constraints from the recent electron EDM experiments can be
greatly relaxed, since $S$ does not acquire a VEV at zero temperature;
thus the mixing of $S$ and the Higgs boson and the $CP$ violation interaction of the top Yukawa is prevented at the tree level;
i.e., the two-loop Barr-Zee contributions to the EDM comes only from the  loop-induced mixing effects.
For example, if one considers $\langle S \rangle = 100$ GeV, then current electron EDM measurements  can exclude the parameter space with $\Lambda < 10$ TeV~\cite{Brod:2013cka}.
This difference can be analytically  understood by loop order estimation. 
In those models with $\langle S \rangle \neq 0$, the $CP$ violation term contributes to electron EDM through the Barr-Zee diagram at the two-loop level.
While in our case with $\langle S \rangle = 0$, this $CP$ violation term can contribute to EDM only at the three-loop level,
 because  the mixing of $H$ and $S$ is induced at the one-loop level.
Thus, in our case the constraints from the EDM are weaker than the collider constraints (discussed in the next section),
which is different from the usual EW baryogenesis case where the EDM constraints are much stronger
than the collide constraints.
Because of  the loop-induced mixing effects,
the two-loop Barr-Zee contribution to EDM is suppressed and
can be expressed as~\cite{Harnik:2012pb,Keus:2017ioh,Brod:2013cka}
\begin{eqnarray}
d_e^{\text{2-loop}} = \frac{e}{3\pi^2} \left( \frac{\alpha_{EW} G_F v}{\sqrt{2}\pi m_t} \right) m_e \left( \frac{v b}{2 \Lambda} \right) \mathcal{O}_{11} \mathcal{O}_{12} \Big[ -g(z_{ts}) + g(z_{th}) \Big],
\end{eqnarray}
with
\begin{eqnarray}
z_{ts} = \frac{m^2_t}{m^2_S} \ , \ z_{th} = \frac{m^2_t}{m^2_H} \ , \
g(z) = \frac{1}{2} z \int^1_0 dx \frac{1}{x(1-x)-z} \log \left(  \frac{x(1-x)}{z}  \right).
\end{eqnarray}
The numerical results are shown in Fig.~\ref{Limit},
where the region below the dotted blue lines is excluded by the
EDM experiments.

We also consider constraints from neutron EDM~\cite{Baker:2006ts,Afach:2015sja,Cirigliano:2016nyn} and mercury EDM~\cite{HgEDM,Yamanaka:2017mef}.
But through our calculation, we find that limits from current neutron and mercury EDM experiments are weaker than electron EDM.
However,  the expected future neutron EDM measurement~\cite{Kumar:2013qya} with a much enhanced precision could
have the capability to detect this type of $CP$ violation.

\begin{figure}[htbp]
\begin{center}
\includegraphics[width=16.0cm]{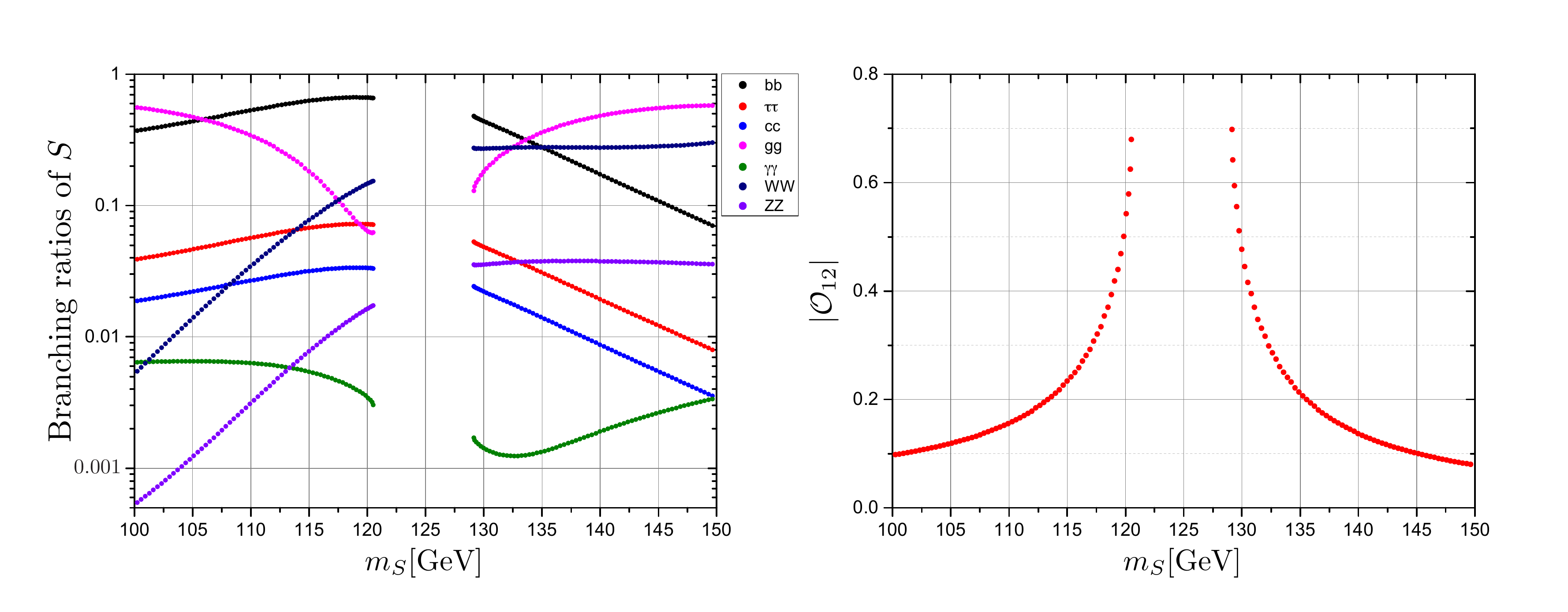}
\caption{Left: Main branching ratios and total decay width of $S$ with different $m_S$. In this plot we set $a$, $b$, and $\Lambda$ as 1, 1, and 1 TeV, respectively.
The mass gap around 125 GeV comes from the $S$-$H$ mixing term $\Delta m^2_{HS} = a\frac{3 m_t^4}{2 \pi^2 \Lambda v}$. The $S$-$H$ mixing term changes the $S$ property hugely when $m_S$ is close to $m_H$.
Right: $S$-$H$ field mixing versus $m_S$ plot. Maximal mixing is obtained when $m_S$ is approaching the boundary of the mass gap.}
\label{S_BR}
\end{center}
\end{figure}

\subsection{Collider direct search and Higgs data}
Production and decay patterns of both the Higgs boson and $S$ particle are modified  by the loop-induced mixing, see Fig.~\ref{S_BR} for an illustration.
In Fig.~\ref{S_BR}, the mass gap around 125 GeV comes from the mass mixing term $\Delta m^2_{HS} = a\frac{3 m_t^4}{2 \pi^2 \Lambda v}$, which is fixed by $\Lambda$ rather than a free parameter.
This feature is shown more clearly in Fig.~\ref{Limit}, where the mass region between black dashed lines is forbidden by this mass mixing term.
Fortran code \texttt{eHDECAY}~\cite{Hdecay1,Hdecay2,Hdecay3,Hdecay4} is used here to do precise calculations.
Figure~\ref{S_BR} shows that the branching ratios of $S$ is quite SM-like near the Higgs mass due to a large mixing with $H$.
While in the region away from 125 GeV, i.e., the region with a smaller mixing, top-loop induced $\gamma\gamma$ and $gg$ channels are enhanced.
Our scenario get constraints from
the SM and non-SM Higgs searches in various channels at LEP, Tevatron, and LHC experiments, and the observed 125 GeV Higgs signal strengths.
We apply cross section upper limits on relevant channels from these collider searches as included in the package \texttt{HiggsBounds}-5~\cite{Hbound1,Hbound2,Hbound3,Hbound4}.
Besides, we use the framework implemented in \texttt{HiggsSignals}-2~\cite{Hsignal} to perform a Higgs data fitting.
Experimental data from $7+8$ TeV ATLAS and CMS combined Higgs measurements~\cite{Khachatryan:2016vau}, and two $13$ TeV Higgs measurements with a higher precision~\cite{Sirunyan:2017exp,Aaboud:2018xdt} are included in the fit.
The Higgs signal strength is required to lie within 2$\sigma$ C.L. of the measured central value. Limits from Higgs data and direct searches are shown in Fig.~\ref{Limit}.
Reading from the figure, the $m_S$ region near 125 GeV is excluded due to the reduced Higgs signal strength through strong mixing between $H$ and $S$,
while in the region with moderate mixing, i.e. the regions away from 125 GeV, limits are mainly from direct resonance searches.
Among them, the most sensitive search channels are the diphoton~\cite{CMS:2015ocq}, and four-lepton~\cite{CMS:2016ilx} final states.
Figure~\ref{Limit} also shows that the limits from the colliders are much stronger than EDM in our scenario.

\subsection{Collider signals in the future}

\begin{figure}
\centering
	\begin{minipage}{.3\textwidth}
		\centering
		\includegraphics[width=0.75\linewidth]{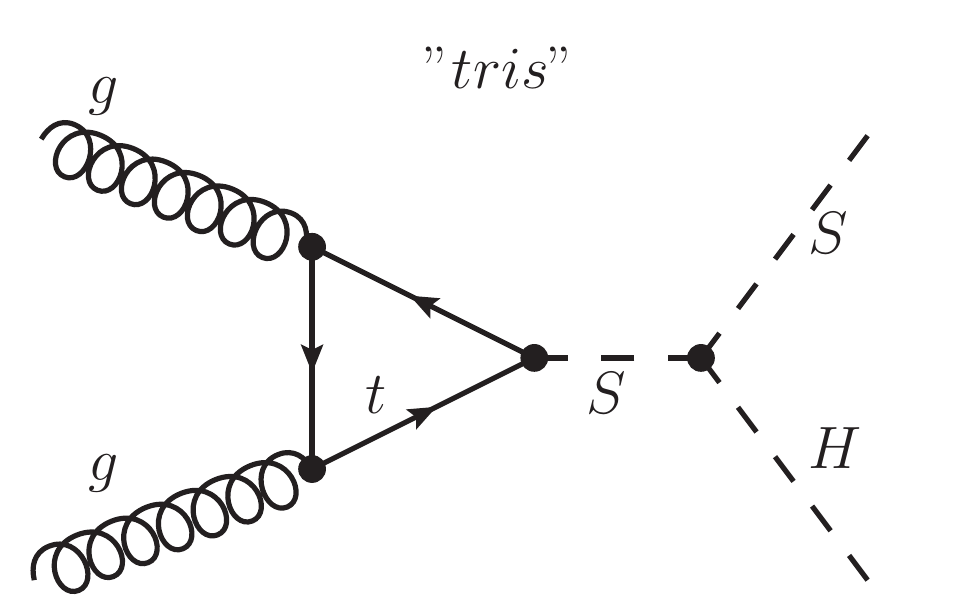}
	\end{minipage}%
	\begin{minipage}{.3\textwidth}
		\centering
		\includegraphics[width=0.75\linewidth]{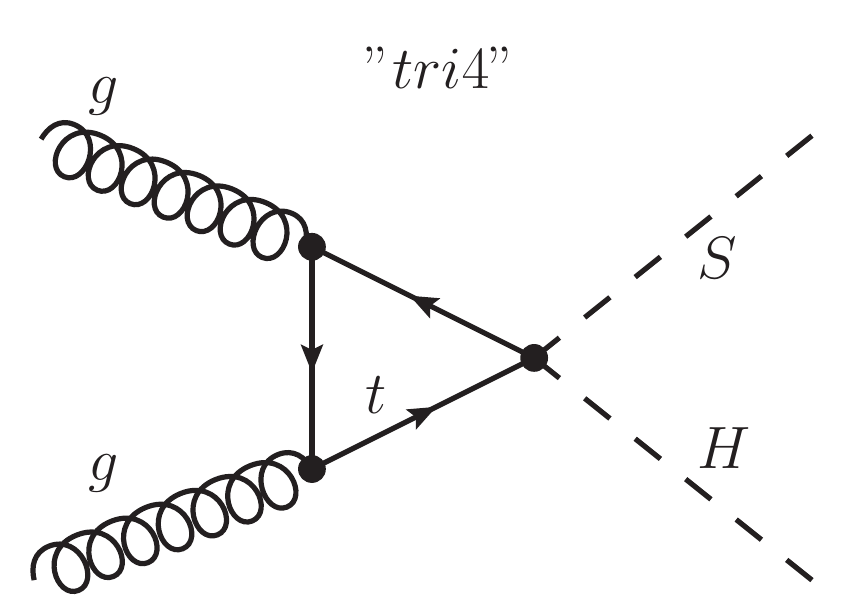}
	\end{minipage}
	\begin{minipage}{.3\textwidth}
		\centering
		\includegraphics[width=0.75\linewidth]{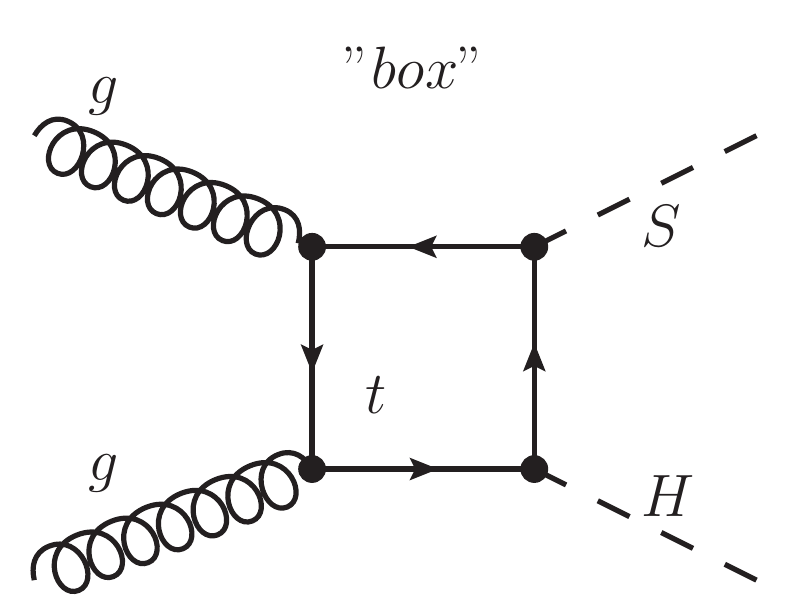}
	\end{minipage}
	\caption{Representative Feynman diagrams that contribute to the $gg\to SH$ process.}
	\label{fig:feynggsh}
\end{figure}

\begin{figure}
\centering
		\includegraphics[width=0.6\linewidth]{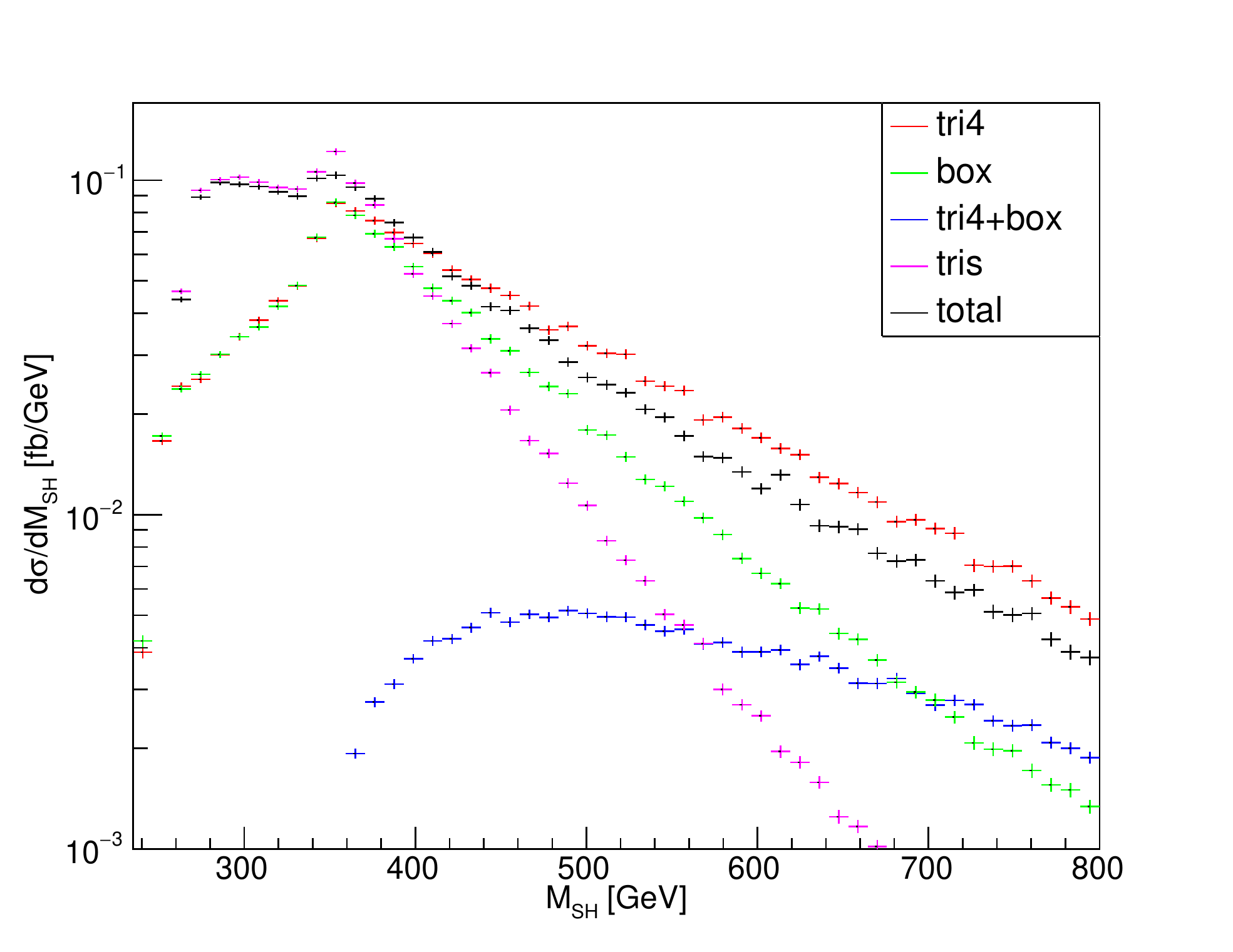}
	\caption{Leading-order differential cross section for the $gg\to SH$ process, with $\kappa = 2$, $m_S = 115$ GeV, $\Lambda = 1$ TeV and $ \sqrt{s} =14$ TeV. The separate contributions from the diagrams are shown in different color schemes.  ``tris"(magenta) represents the cross section considering only the first type of diagrams as in Fig.\ref{fig:feynggsh}, ``tri4"(red) represents the second, and ``box"(green) represents the last. The blue curve shows the cross section including the ``tri4" and the ``box" contributions. The black curve is the total cross section including all diagrams and their interference, which is dominated by the ``tris", or $\kappa$ term at a low energy scale, and by the dimension-five $\eta$ term and interference at a high energy scale.}
	\label{fig:ggsh_mzh}
\end{figure}

There are several channels in our model that may produce observable signals at high luminosity LHC,
for example: $pp \to S \to jj$, $pp \to S \to \gamma\gamma$, $pp \to S \to Z Z^{\ast} \to l^+l^-l^+l^-$, and $pp \to SH$.
The light di-jet resonance search suffers from a huge QCD background~\cite{Sirunyan:2017nvi} and remains difficult even at a future LHC run.
Due to a much less background, previous diphoton and four-lepton search results, as shown earlier, already excluded some parameter space of our model.
So di-photon and 4-lepton channels would continue to exclude parameter space or give the first hint of signals as the LHC continues accumulating data.
In Table~\ref{lhc_gz} we give the production cross sections times branching ratios at 14 TeV LHC of these two channels for the two benchmark points.
 A concrete analysis relies on detailed simulation and dedicated final state studies, which is beyond the scope of the current paper, and could be interesting future work.
The $pp \to SH$ process is mostly through the one-loop $gg\to SH$ contribution, and an exact calculation at the leading order is performed. There are three types of Feynman diagrams as shown in Fig.\ref{fig:feynggsh}. The second (tri4) and third (box) diagrams are proportional to the contribution of the dimension-five effective operator, and thus interfere destructively according to the low-energy theorem \cite{Kniehl:1995tn}. Their contributions nearly cancel out at low-energy scale, just above the $m_S+m_H$ threshold. The first diagram (tris), however, is proportional to $\kappa$, and contributes dominantly when $\kappa$ becomes large.
The leading-order total cross section of $pp \to SH$ is around 25 fb with $\kappa = 2$, $m_S = 115$ GeV, $\Lambda = 1$ TeV, and $ \sqrt{s} =14$ TeV, and roughly scales with $\kappa^2$ for even larger $\kappa$ values.
We illustrate the separate contributions to the leading order differential cross section as a function of $m_{SH}$ from the different diagrams in Fig.~\ref{fig:ggsh_mzh}. As seen in the figure, the total contribution is indeed dominated by the ``tris", or $\kappa$ term at low-energy scale, and dominated by the ``tri4+box", or the dimension-five term proportional to $\eta$ at high energy scale. Thus by probing this process, we obtain complementary information on the model parameters compared to the diphoton and four-lepton search.
Multiplied by a $k$ factor of around two for typical $gg$ to scalar(s) processes, this $gg\to SH$ process becomes comparable to or even larger than the SM $pp \to HH$ total cross section, which is about 40 fb at 14 TeV. In our scenario, the $S$ decays dominantly to a pair of gluons and by a small fraction to a pair of photons. A study that is similar to the di-Higgs search at the high luminosity LHC, while with one scalar at a different mass, in the $\gamma\gamma b\bar b$ and $jj b\bar b$ final states, becomes another interesting future work.
The $pp\to SH$ study would benefit from a future hadron collider with a higher center of mass energy, for example at a 27 TeV HE-LHC and a 100 TeV FCC-hh, SPPC. Very similar to the study of di-Higgs production,  the cross section of the $gg\to SH$ increases from 25 to 92  and 770 fb at 27  and 100 TeV center of mass energy, respectively, with our leading-order calculation.

Note here that the scalar $S$ is larger than half the Higgs mass in our benchmark scenarios and cannot be produced or probed through Higgs decay; the $\frac{1}{2}\kappa S^2\Phi^2$ term with large $\kappa$ could as well be indirectly probed at the off-shell Higgs region, for example, as discussed in Ref.~\cite{Goncalves:2017iub}.

\begin{table}[h!]
\centering
\caption{Production cross sections of $S$ times branching ratios at 14 TeV LHC, with $\Lambda$ = 1 TeV.}
\begin{tabular}{c|c|ccccc}
  \hline
  \hline
$m_S$[GeV]               &  $\sigma(pp\to S) \times BR(S\to \gamma\gamma)$              & $\sigma(pp\to S) \times BR(S\to ZZ^{\ast})$    \\ \hline
115              & 37.73 fb                      &  54.69 fb    \\   \hline
135              & 18.38 fb                 & 520.60 fb  \\  \hline   \hline
\end{tabular}
\label{lhc_gz}
\end{table}

\begin{figure}[htbp]
\begin{center}
\includegraphics[width=16.0cm]{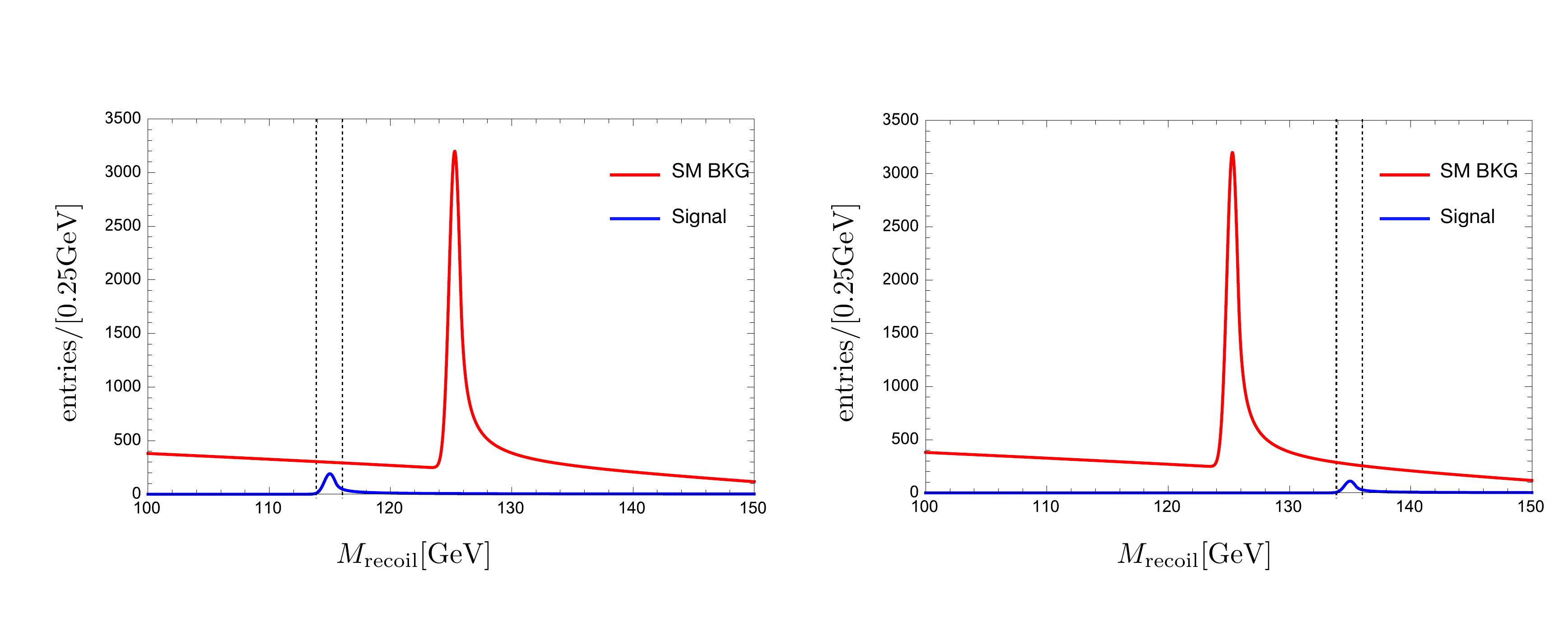}
\caption{Left: $\mu\mu$ recoil mass distribution for the SM background and signal, with $\Lambda =$ 1 TeV and $m_S =$ 115 GeV.
Right: $\mu\mu$ recoil mass distribution for the SM background and signal, with $\Lambda =$ 1 TeV and $m_S =$ 135 GeV.
Vertical dotted black lines represent the mass window we choose. Luminosity is taken at 5 ${ab}^{-1}$ following the CEPC report.
The $y$ axis represents the number of events per bin, which is taken to be 0.25 GeV.}
\label{CEPC}
\end{center}
\end{figure}

Meanwhile, collider signal searches at future electron-positron colliders like the CEPC are much more clean and promising.
Here we do a simple analysis by applying the recoiled $\mu\mu$ mass distribution at a 5 ${ab}^{-1}$ luminosity CEPC to estimate our sensitivity.
The SM Higgs boson and other SM background distributions are described by a Crystal Ball function and third-order Chebychev polynomial function respectively.
Parameters are fixed by fitting with the CEPC group report~\cite{Chen:2016zpw}.
The signal is a scalar-strahlung process $e^+ e^- \to Z^\ast \to ZS$, with a total cross section~\cite{Mo:2015mza}
\begin{eqnarray}
\sigma(e^+ e^- \to ZS) = \frac{G_F ^2 m^4_Z}{96 \pi s} (v^2_e + a^2_e) |\mathcal{O}_{12}|^2 \sqrt{\tilde{\lambda}}~ \frac{\tilde{\lambda} + 12m^2_Z/s}{(1-m^2_Z/s)^2}~.
\end{eqnarray}
Here $v_e = -1 + 4s_w^2$, $a_e = -1$, and $\tilde{\lambda} = (s^2+m^4_Z+m^4_S-2sm^2_Z-2sm^2_S-2m^2_Sm^2_Z)/s^2$ where $\sqrt{s} = 250$ GeV, $s_w$ is sine of the Weinberg angle.
The shape of the signal peak is estimated and obtained by a rescaling and shifting from the fitted SM Higgs shape. Figure \ref{CEPC} shows the recoil mass distribution.
Then we count the number of SM background and signal events in the [$m_S - 1$GeV, $m_S + 1$GeV] mass window, noted as $\mathcal{B}$ and $\mathcal{S}$ respectively.
So the significance can be written as $\mathcal{S}/\sqrt{\mathcal{B}+\epsilon^2 \mathcal{B}^2}$, with $\epsilon = 1.0\%$ being the dominant systematic uncertainty.
The region with $\mathcal{S}/\sqrt{\mathcal{B}+\epsilon^2 \mathcal{B}^2} > 5$ can be observed at \text{5 ${ab}^{-1}$} CEPC with a significance higher than 5$\sigma$, and  the curve is shown as well in Fig.~\ref{Limit}.
It is clear from Fig.~\ref{Limit} that there is a large currently allowed parameter space that can be covered by High Luminosity LHC or CEPC. We are especially sensitive to regions with $m_S$ closer to 125 GeV, which corresponds to an increasing $S$-$H$ mixing.

In addition,
$S$-$H$ mixing could also be detected through a potentially visible deviation of $\sigma(e^+ e^- \to HZ)$ measurement, which can be an indirect signal of our model~\cite{Cao:2017oez}.
Furthermore, wave function renormalization of the Higgs field which comes from $ \frac{1}{2} \kappa S^2 (\Phi^\dagger \Phi)$ reduces $\sigma(e^+ e^- \to HZ)$ by a global rescaling factor:
\begin{eqnarray}
\mathcal{Z} = 1 - \frac{\kappa^2 v^2}{32 \pi^2 m_H^2} \left( \frac{4m_S^2}{m_H^2} \frac{1}{\sqrt{\frac{4m_S^2}{m_H^2}-1}  } \arctan \frac{1}{\sqrt{\frac{4m_S^2}{m_H^2}-1}  } -1       \right)~.
\end{eqnarray}
As a result, the total cross section $\sigma(e^+ e^- \to HZ)$ will be rescaled by a factor of $|\mathcal{O}_{22}|^2 \mathcal{Z}$.
Quoting from the proposed precision of CEPC with 5 ab$^{-1}$ data,  it is capable to measure the inclusive $HZ$ cross section to about $1.0\%$ sensitivity.
In Fig.~\ref{Limit} we draw contour lines for different ratio $\frac{\sigma(HZ)}{\sigma_{SM}(HZ)}$.
Unlike the nearly symmetric shape the direct search lines, $\sigma(HZ)$ shows a larger deviation in the lighter $m_S$ region.
This effect comes from the Higgs field wave function renormalization, which is more sensitive to a lighter $m_S$.
This indirect detection method shows good sensitivity, and gives complementary information on the model parameters in addition to our direct search.

\section{Gravitational wave signals and their correlation with collider signals}
\label{subsec:GW}
The key point to predict the phase transition GW signal is to calculate the two parameters $\alpha$ and $\tilde{\beta}$
from the finite temperature effective potential in Eq.~(\ref{fullpotential}) using the method described in Sec.~\ref{sec:EWPT}.
The two parameters are related to the phase transition strength and
the inverse of the time duration, respectively.
The GWs also depend on the energy efficiency factors $\lambda_{i}$ (i=col, turb, sw, denoting bubble collision, turbulence, and sound waves, respectively) and bubble wall velocity $v_b$.
For the GW spectrum from bubble collisions, we use the formulas from
the envelope approximations~\cite{Huber:2008hg,Jinno:2016vai}:
\begin{align}
  \Omega_{\rm col} (f) h^2
  \simeq 1.67 \times 10^{-5}  \times\left(\frac{0.11v_b^3}{0.42+v_b^2}\right)
  \tilde{\beta}^{-2}
  \left(\frac{\lambda_{\rm col} \alpha}{1+\alpha}\right)^2
  \left(\frac{100}{g_*^{}(T_N^{})}\right)^{1/3}
  \frac{3.8 (f/\tilde{f}_{\rm col})^{2.8}}{1+2.8 (f/\tilde{f}_{\rm col})^{3.8}},  \nonumber
\end{align}
at the peak frequency
\begin{align}
  \tilde{f}_{\rm col} \simeq
  1.65 \times 10^{-5}~\Hz \times \left(\frac{0.62}{1.8-0.1v_b+v_b^2}\right)
  \tilde{\beta}
  \left(\frac{T_N}{100~\GeV}\right)\left(\frac{g_*^{}(T_N^{})}{100}\right)^{1/6}.
\end{align}
The efficiency factor  $\lambda_{\rm col}$ is a function of
$\alpha$ and $v_b$, and
we use the results for the deflagration case as obtained in Ref.~\cite{Espinosa:2010hh}.
As for a GW spectrum from sound waves,
numerical simulations give~\cite{Hindmarsh:2013xza,Caprini:2015zlo}
\begin{align}
  \Omega_{\rm sw} (f) h^2
  \simeq    2.65 \times 10^{-6} v_b
  \tilde{\beta}^{-1}
  \left(\frac{\lambda_{\rm sw} \alpha}{1+\alpha}\right)^2
  \left(\frac{100}{g_*^{}(T_N^{})}\right)^{1/3}\times
    (f/\tilde{f}_{\rm sw})^3
  \left(\frac{7}{4+3(f/\tilde{f}_{\rm sw})^2}\right)^{7/2}   \nonumber
  \end{align}
with the peak frequency
\begin{align}
  \tilde{f}_{\rm sw} \simeq 1.9 \times 10^{-5}~\Hz \frac{1}{v_b}
  \tilde{\beta}
  \left(\frac{T_N}{100~\GeV}\right)
  \left(\frac{g_*^{}(T_N^{})}{100}\right)^{1/6}.
\end{align}
The turbulence contribution to the GW spectrum
is~\cite{Binetruy:2012ze,Caprini:2009yp}
\begin{align}
  \Omega_{\rm turb} (f) h^2
  \simeq  3.35 \times 10^{-4} v_b
  \tilde{\beta}^{-1}
  \left(\frac{\lambda_{\rm turb} \alpha}{1+\alpha}\right)^{3/2}
\left(\frac{100}{g_*^{}(T_N^{})}\right)^{1/3} \times
  \frac{(f/\tilde{f}_{\rm turb})^3 }{(1+ f/\tilde{f}_{\rm turb})^{11/3}
  (1+8 \pi f/ \mathcal{H}_0)}       \nonumber
\end{align}
with the peak frequency
\begin{align}
  \tilde{f}_{\rm turb} \simeq 2.7 \times 10^{-5}~\Hz \frac{1}{v_b}
  \tilde{\beta}
  \left(\frac{T_N}{100~\GeV}\right)
  \left(\frac{g_*^{}(T_N^{})}{100}\right)^{1/6}
\end{align}
and
\begin{align}
  \mathcal{H}_0=1.65 \times 10^{-5}~\Hz
  \left(\frac{T_N}{100~\GeV}\right)
  \left(\frac{g_*^{}(T_N^{})}{100}\right)^{1/6}.
\end{align}

We now show our numerical results of the total GW spectrum
from the three contributions in the concerned scenario with the benchmark parameter sets.
\begin{figure}[ht]
\begin{center}
\includegraphics[width=0.5\textwidth,clip]{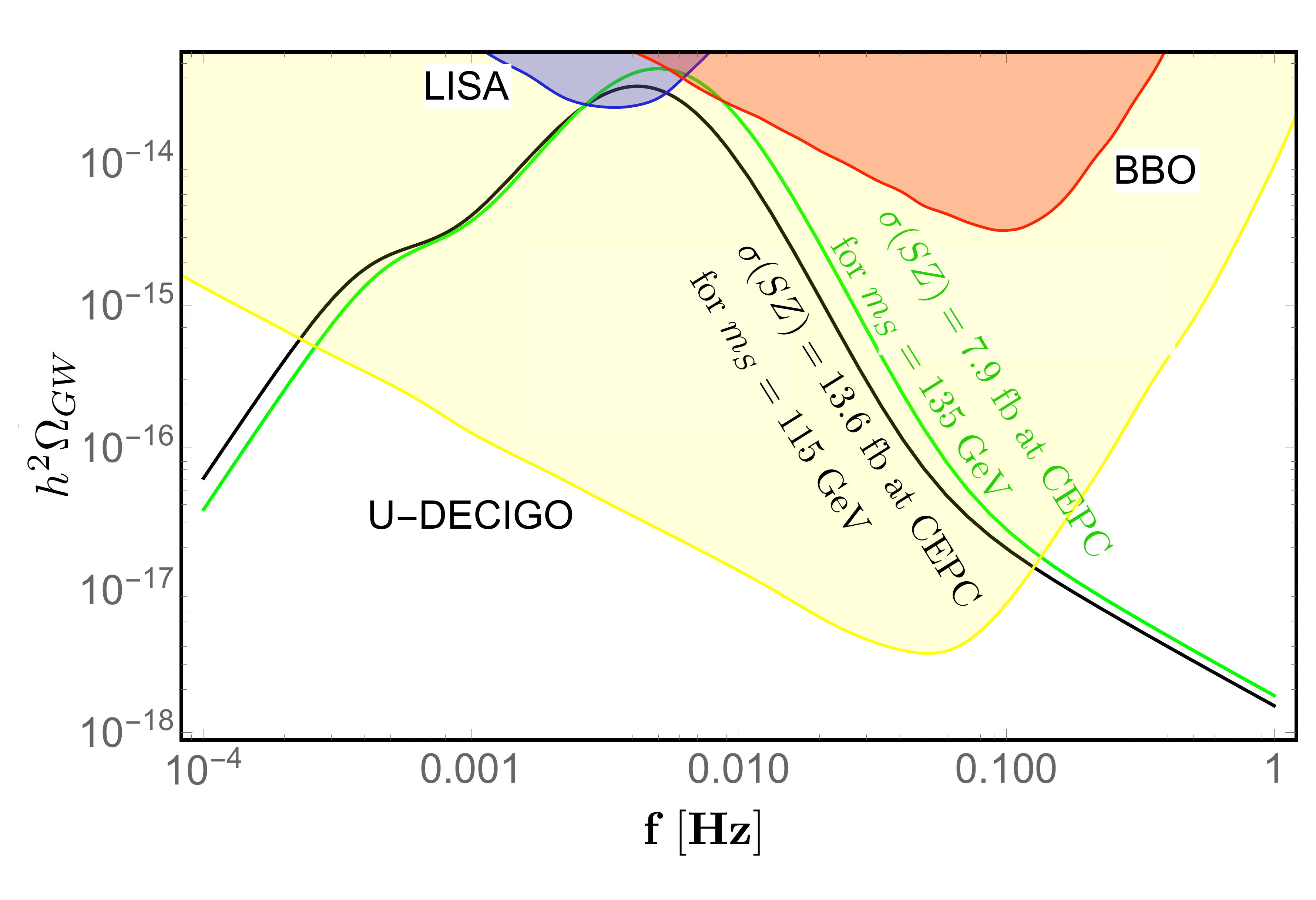}\\
\caption{ The correlation between the GW spectrum and the associated collider signals
for the benchmark sets with $\kappa=2$ and $\Lambda=1$ TeV.
The colored regions depict the expected sensitivities from the future GW experiments LISA, BBO and U-DECIGO, respectively.
The black line represents the phase transition GW spectrum for the benchmark sets at $m_S=115$ GeV, which is related to the
detectable lepton collider signal with a cross section $\sigma(SZ)=13.6$ fb at CEPC . The green line represents the case for another benchmark set at $m_S=135$ GeV.}\label{gws}
\end{center}
\end{figure}
From Fig.~\ref{gws},
we can see that the GWs produced in this EW baryogenesis scenario can be detected marginally by
LISA, BBO and certainly by U-DECIGO.
We also show the corresponding CEPC cross sections as a double test on this
scenario, and vice versa.
For example taking benchmark set I, the GW spectrum is represented by the black line in Fig.~\ref{gws},
which can be detected by LISA and U-DECIGO. 
The black line also corresponds to $0.9339\sigma_{SM}(HZ)$  of the HZ cross section for  $e^+e^- \to HZ$ process 
and  115 GeV  recoil mass with 13.6 fb cross section for the $e^+e^- \to SZ$ process,
which has a 5$\sigma$ discovery potential with 5 $ab^{-1}$ luminosity at CEPC.  
Other lepton colliders are also capable to detect this collider signals, such as ILC and FCC-ee.
The observation of GWs with a several mHz peak frequency at LISA and the observation of the 115 GeV recoil mass at CEPC
are related by this EW baryogenesis scenario.
We can see that the
 future lepton collider and GW detector can make a double test on the scenario~\cite{Huang:2016odd,Huang:2015izx,Huang:2017rzf,Huang:2017kzu}.

\section{Conclusion}\label{sec:conclusion}

We have studied the collider search and GW detection of the EW baryogenesis scenario with a dynamical source of $CP$ violation realized by a two-step phase transition.
The VEV of a new scalar field $ S $ evolves with the two-step phase transition,  and provides both the SFOPT and sufficient $CP$ violation at the early universe.
At the current time,  the VEV of $S$ becomes zero at the tree level, which makes it easy to evade the
severe EDM constraints.  Nevertheless, the loop-induced mixing between the scalars $S$ and $H$ can produce abundant collider signals.
We have shown the possible  collider signals at future collider experiments, especially at the lepton colliders.
Meanwhile, collider signals and GW surveys could  cross-check  this EW baryogenesis  scenario.
As a by-product, the discussion here suggests potentially interesting collider signals for additional generic light scalar searches near the Higgs mass.
The analysis in this work may help to  understand the origin of $CP$ violation and EW baryogenesis,
furthering the connection between cosmology and particle physics.
More systematical study is left to our future study.
\begin{acknowledgments}
We deeply appreciate Ryusuke Jinno's comments concerning gravitation wave generation from phase transition, and Eibun Senaha's helpful discussion on the $CP$ violation source in EW baryogenesis.
We also thank Takumi Kuwahara for his useful suggestion on accounting for the EDM contribution, and Taehyun Jung on wave function renormalization. This work is supported by IBS under the project code IBS-R018-D1.
\end{acknowledgments}

\medspace
\medspace
\medspace
\medspace
\medspace
\medspace
\medspace
\medspace

\noindent {\bf Appendix: Effective Lagrangian calculation through covariant derivative expansion}

CDE is a convenient method to calculate one-loop effective Lagrangian~\cite{Gaillard:1985uh,Cheyette:1987qz,Henning:2014wua}.
In this appendix we use CDE to calculate several most important operators in our work.
Operators we calculated here are those connecting gluon or photon pairs to scalars and induced by the top-quark loop, which are most relevant to the phenomenology we want to study at hadron colliders.
In order to make things clear and easy to check, we write down our calculation procedure in detail.
For notation and convention, we follow Ref.~\cite{Henning:2014wua}.

The particle being integrated out here is the  top quark, so the corresponding one-loop contribution to the effective action is:
\begin{eqnarray}
S_\text{eff,1-loop} = -i \text{Tr}\log \left( P\!\!\!\!/ - m_t - M \right),
\end{eqnarray}
with $P_{\mu} \equiv iD_{\mu} $, and $ D_{\mu} = \partial_{\mu} - i\frac{2}{3}eA_{\mu} - ig_3 T^a G^a_{\mu}$.
$M$ is the bilinear coefficient of the top-quark field:
\begin{eqnarray}
M = aS \left( \frac{m_t}{\Lambda} + \frac{m_t H}{\Lambda v} \right) + i b S \gamma_5 \left(\frac{m_t}{\Lambda} + \frac{m_t H}{\Lambda v} \right) + \frac{m_t H}{v} .
\end{eqnarray}
$S_\text{eff,1-loop}$ can thus be rewritten as
\begin{eqnarray}
S_\text{eff,1-loop} &=& -\frac{i}{2} \text{Tr}\log \left( - P^2 + m^2_t - \frac{i}{2}\sigma^{\mu\nu} G^{\prime}_{\mu\nu} + 2m_t M + M^2 + [P\!\!\!\!/ , M]  \right)\\\nonumber
                             &\equiv& -\frac{i}{2} \text{Tr}\log \left( - P^2 + m^2_t + U \right) ,
\end{eqnarray}
where $G^{\prime}_{\mu\nu} = [D_{\mu} , D_{\nu}]$,and $\sigma^{\mu\nu} = \frac{i}{2}[\gamma^{\mu},\gamma^{\nu}]$.
After separating the covariant derivatives and the loop momentum, one-loop effective Lagrangian can be written as
\begin{eqnarray}
\Delta \mathcal{L}_\text{eff,1-loop} = \frac{i}{2} \int dq \int dm^2 \text{tr} \frac{1}{\Delta^{-1}\left[1 - \Delta\left(-\left\{q^{\mu},\widetilde{G}_{\nu\mu}\right\} \partial^{\nu} -
\widetilde{G}_{\mu\sigma}\widetilde{G}_{\nu}^{\  \sigma} \partial^{\mu}\partial^{\nu} + \widetilde{U} \right)\right] }.
\end{eqnarray}
Here $dq \equiv d^4q/(2\pi)^4$, $\Delta \equiv 1/(q^2 - m^2)$(for our case, $m = m_t$), and:
\begin{eqnarray}
\widetilde{G}_{\mu\nu} = \sum^{\infty}_{n=0}\frac{n+1}{(n+2)!}   \left[P_{\alpha_1},\left[P_{\alpha_2},\left[...\left[P_{\alpha_n},\left[D_{\mu}, D_{\nu}\right]\right]\right]\right]\right]
\frac{\partial^n}{\partial q_{\alpha_1}\partial q_{\alpha_2}...\partial q_{\alpha_n}} , \\
 \widetilde{U} = \sum^{\infty}_{n=0}\frac{1}{n!}   \left[P_{\alpha_1},\left[P_{\alpha_2},\left[...\left[P_{\alpha_n},U\right]\right]\right]\right]
\frac{\partial^n}{\partial q_{\alpha_1}\partial q_{\alpha_2}...\partial q_{\alpha_n}}.
\end{eqnarray}
The trace ``tr" here  acts only on indices like the spin, generation, and flavor, but not the momentum.
Then $\Delta \mathcal{L}_\text{eff,1-loop}$ can be expanded by a series of integral:
\begin{eqnarray}
& &  \Delta \mathcal{L}_\text{eff,1-loop} = \frac{i}{2} \sum^{\infty}_{n=0} \mathcal{I}_n , \label{equ1} \\
\mathcal{I}_n &\equiv & \text{tr} \int dq dm^2 \left[\Delta\left(-\left\{q^{\mu},\widetilde{G}_{\nu\mu}\right\} \partial^{\nu} -
\widetilde{G}_{\mu\sigma}\widetilde{G}_{\nu}^{\  \sigma} \partial^{\mu}\partial^{\nu} + \widetilde{U} \right)\right]^n \Delta.
\label{equ2}
\end{eqnarray}
In most cases, one does not  need all the terms in Eqs. (\ref{equ1}) and (\ref{equ2}), and only the first few terms are important. In our case, the relevant terms we need to calculate contain at least two vector field strengths $G_{\mu\nu}^{\prime}$ and at least one scalar $S$ or $H$, and do not contain the derivatives of these fields. So the sum of all the relevant terms, after loop momentum integral, is
\begin{eqnarray}
\Delta \mathcal{L}_\text{eff} = & & -\frac{1}{2}\frac{1}{(4\pi)^2}\text{tr} \bigg\{ \frac{1}{m_t^2}\left(-\frac{1}{6}U^3-\frac{1}{12}UG^{\prime}_{\mu\nu}G^{\prime\mu\nu}   \right)     \\\nonumber
 & &+ \frac{1}{m_t^4}\left( \frac{1}{24}U^4 + \frac{1}{24} \left(U^2 G^{\prime}_{\mu\nu}G^{\prime\mu\nu} \right)  \right)   \bigg\} .
\end{eqnarray}
Then we calculate the trace and express the effective Lagrangian by $S$, $H$, gluon field strength $G^a_{\mu\nu}$, and photon field strength $F_{\mu\nu}$.
In order to make the calculation clear and get a concise expression, we introduce some useful notations.

The trace of two covariant derivative commuters is
\begin{eqnarray}
\text{tr}\left( G^{\prime}_{\mu\nu} G^{\prime}_{\alpha\beta} \right) = -\frac{4}{3}e^2 F_{\mu\nu} F_{\alpha\beta} - \frac{1}{2}g^2_3 G^a_{\mu\nu} G^a_{\alpha\beta}~.
\end{eqnarray}
The trace can be divided into two parts, with or without a $\gamma^5$:
\begin{eqnarray}
\text{tr}\left(\sigma^{\mu\nu}G^{\prime}_{\mu\nu} \sigma^{\alpha\beta}G^{\prime}_{\alpha\beta}  \right)
= \text{tr} \left( \sigma^{\mu\nu} \sigma^{\alpha\beta} \right) \text{tr} \left( G^{\prime}_{\mu\nu} G^{\prime}_{\alpha\beta} \right)
\end{eqnarray}
\begin{eqnarray}
\text{tr}\left(\sigma^{\mu\nu}G^{\prime}_{\mu\nu} \sigma^{\alpha\beta}G^{\prime}_{\alpha\beta} \gamma_5 \right)
= \text{tr} \left( \sigma^{\mu\nu} \sigma^{\alpha\beta} \gamma_5  \right) \text{tr} \left( G^{\prime}_{\mu\nu} G^{\prime}_{\alpha\beta} \right)~.
\end{eqnarray}
and then, by using the identities
\begin{eqnarray}
\text{tr}\left( [\gamma^{\mu},\gamma^{\nu}] [\gamma^{\alpha},\gamma^{\beta}] \right) = 16(g^{\mu\beta}g^{\nu\alpha} - g^{\mu\alpha}g^{\nu\beta}),
\end{eqnarray}
\begin{eqnarray}
\text{tr}\left( [\gamma^{\mu},\gamma^{\nu}] [\gamma^{\alpha},\gamma^{\beta}] \gamma_5 \right) = - i 16 \epsilon^{\mu\nu\alpha\beta},
\end{eqnarray}
we get
\begin{eqnarray}
\text{tr}\left(\sigma^{\mu\nu}G^{\prime}_{\mu\nu} \sigma^{\alpha\beta}G^{\prime}_{\alpha\beta}  \right)
=   8 g^{\mu\alpha}g^{\nu\beta} \text{tr} \left( G^{\prime}_{\mu\nu} G^{\prime}_{\alpha\beta} \right)
\end{eqnarray}
\begin{eqnarray}
\text{tr}\left(\sigma^{\mu\nu}G^{\prime}_{\mu\nu} \sigma^{\alpha\beta}G^{\prime}_{\alpha\beta} \gamma_5 \right)
= i 4 \epsilon^{\mu\nu\alpha\beta} \text{tr} \left( G^{\prime}_{\mu\nu} G^{\prime}_{\alpha\beta} \right).
\end{eqnarray}
Now we define
\begin{eqnarray}
\mathcal{I}_A \equiv & & 2m_t \left(aS\left( \frac{m_t}{\Lambda} + \frac{m_t H}{\Lambda v} \right) + \frac{m_t H}{v} \right) \\\nonumber
       & &+ \left(aS\left( \frac{m_t}{\Lambda} + \frac{m_t H}{\Lambda v} \right) + \frac{m_t H}{v} \right)^2 - b^2S^2\left( \frac{m_t}{\Lambda} + \frac{m_t H}{\Lambda v} \right)^2~,
\end{eqnarray}
and
\begin{eqnarray}
\mathcal{I}_B \equiv & & 2m_t bS\left( \frac{m_t}{\Lambda} + \frac{m_t H}{\Lambda v} \right) \\\nonumber
        & & + 2bS\left( \frac{m_t}{\Lambda} + \frac{m_t H}{\Lambda v} \right)\left(aS\left( \frac{m_t}{\Lambda} + \frac{m_t H}{\Lambda v} \right) + \frac{m_t H}{v} \right).
\end{eqnarray}
Using these replacements, $U$ can be rewritten in the simple form
\begin{eqnarray}
U = -\frac{i}{2} \sigma^{\mu\nu} G^{\prime}_{\mu\nu} + \mathbf{1} \mathcal{I}_A + i \gamma_5 \mathcal{I}_B + [P\!\!\!\!/ , M]~.
\end{eqnarray}
Then we can easily express those traces (here we show only those relevant terms):
\begin{eqnarray}
\text{tr}\left( U^3 \right) \supset -\frac{3}{4}\left( 8\mathcal{I}_A g^{\mu\alpha}g^{\nu\beta} -4\mathcal{I}_B \epsilon^{\mu\nu\alpha\beta}   \right) \text{tr}\left( G^{\prime}_{\mu\nu}G^{\prime}_{\alpha\beta} \right)      \\
\text{tr}\left( U G^{\prime}_{\mu\nu}G^{\prime\mu\nu} \right) \supset 4 \mathcal{I}_A \text{tr}\left( G^{\prime}_{\mu\nu}G^{\prime\mu\nu} \right)      \\
\text{tr}\left( U^4 \right) \supset -\frac{3}{2} \left( 8 \left( \mathcal{I}_A^2 -\mathcal{I}_B^2 \right) g^{\mu\alpha}g^{\nu\beta} - 8 \mathcal{I}_A \mathcal{I}_B \epsilon^{\mu\nu\alpha\beta} \right) \text{tr}\left( G^{\prime}_{\mu\nu}G^{\prime}_{\alpha\beta} \right)      \\
\text{tr}\left( U^2 G^{\prime}_{\mu\nu}G^{\prime\mu\nu} \right) \supset 4\left( \mathcal{I}_A^2 - \mathcal{I}_B^2 \right) \text{tr}\left( G^{\prime}_{\mu\nu}G^{\prime\mu\nu} \right)~.
\end{eqnarray}
Then the top-quark loop-induced effective coupling between vector pairs and the scalars can be obtained as
\begin{eqnarray}
\mathcal{L}_{SVV} &=& \frac{1}{3} \frac{1}{m^2_t} \mathcal{I}_A \left( \frac{4}{3} \frac{\alpha_{EW}}{4 \pi} F_{\mu\nu}F^{\mu\nu} + \frac{1}{2} \frac{\alpha_S}{4 \pi} G^a_{\mu\nu}G^{a\mu\nu} \right) \\\nonumber
 &-& \frac{1}{2} \frac{1}{m^2_t} \mathcal{I}_B \left( \frac{4}{3} \frac{\alpha_{EW}}{4 \pi} F_{\mu\nu}\tilde{F}^{\mu\nu} + \frac{1}{2} \frac{\alpha_S}{4 \pi} G^a_{\mu\nu}\tilde{G}^{a\mu\nu} \right) \\\nonumber
&-&  \frac{1}{6} \frac{1}{m^4_t} (\mathcal{I}_A^2 - \mathcal{I}_B^2) \left( \frac{4}{3} \frac{\alpha_{EW}}{4 \pi} F_{\mu\nu}F^{\mu\nu} + \frac{1}{2} \frac{\alpha_S}{4 \pi} G^a_{\mu\nu}G^{a\mu\nu} \right) \\\nonumber
 &+& \frac{1}{2} \frac{1}{m^4_t} \mathcal{I}_A \mathcal{I}_B \left( \frac{4}{3} \frac{\alpha_{EW}}{4 \pi} F_{\mu\nu}\tilde{F}^{\mu\nu} + \frac{1}{2} \frac{\alpha_S}{4 \pi} G^a_{\mu\nu}\tilde{G}^{a\mu\nu} \right)~.
\end{eqnarray}
The calculation of scalar mass corrections is much easier.
Scalar mass corrections come from terms with no field derivatives:
\begin{eqnarray}
\mathcal{L}_{\Delta m^2} \subset -\frac{1}{2} \frac{1}{(4\pi)^2} \text{tr} \bigg\{-m_t^2 \left( \log\frac{m_t^2}{\mu_R^2} - 1 \right) U -\frac{1}{2} \log\frac{m_t^2}{\mu_R^2} U^2 \bigg\}.
\end{eqnarray}
Expanding $U$ and setting renormalisation scale $\mu_R$ as $m_t$, we obtain
\begin{eqnarray}
\mathcal{L}_{\Delta m^2} = -\frac{3}{8\pi^2}m_t^2 \left( \frac{m_t^2}{v^2} H^2 + 4a\frac{m_t^2}{\Lambda v}SH + (a^2 - b^2)\frac{m_t^2}{\Lambda^2} S^2 \right)  .
\end{eqnarray}


\end{document}